\documentclass[11pt]{article}

\usepackage[letterpaper, margin=1in]{geometry}
\usepackage[T1]{fontenc}
\usepackage{lmodern}
\usepackage{amsmath, amssymb, mathtools}
\usepackage{graphicx}
\usepackage{caption}
\usepackage{subcaption}
\usepackage{microtype}
\usepackage{enumitem}
\usepackage[colorlinks=true, allcolors=blue]{hyperref}
\usepackage{cite}

\newcommand{\Ag}{\mathrm{Ag}}
\newcommand{\Cl}{\mathrm{Cl}}
\newcommand{\Kp}{\mathrm{K}}

\newcommand{\avg}[1]{\langle #1 \rangle}
\newcommand{\rmd}{\mathrm{d}}
\graphicspath{{figures/}}

\newlist{parts}{enumerate}{2}
\setlist[parts]{label=(\alph*),leftmargin=2em,labelsep=0.5em,
                topsep=4pt,itemsep=2pt}

\title{A class and home problem on electrolyte transport: \\
       constant electric field implies electroneutrality, but \\
       electroneutrality does not imply a constant electric field}

\author{Ankur Gupta\thanks{Department of Chemical and Biological Engineering,
Department of Applied Mathematics, and Materials Science and Engineering Program,
University of Colorado Boulder, Boulder, CO 80309, USA.
\texttt{ankur.gupta@colorado.edu}}}

\date{\today}

\begin{document}
\maketitle

\begin{abstract}
We present a class and home problem in graduate transport phenomena and
electrochemical engineering that clarifies a common misconception:
electroneutrality implies the electric field is constant. Starting with
one-dimensional Poisson--Nernst--Planck equations for a silver electroplating
cell, students obtain concentration and potential profiles. A companion
home problem with added background electrolyte introduces a new
dimensionless ratio and admits a closed-form solution. Students conclude
that electroneutrality is necessary but not sufficient for a constant
electric field.
\end{abstract}

\section{Introduction}

Electrochemical engineering has shifted from a specialized topic to a
routine part of the chemical engineering curriculum. Many programs now
offer at least one course in the area, typically as a senior elective
that is cross-listed with graduate students. The course serves students
who go on to work in batteries, electrolyzers, fuel cells,
electrodeposition, and electrochemical separations, and its visibility
has grown alongside the electrification of the chemical
industry~\cite{schiffer2017}. In addition to electrochemical engineering,
graduate transport phenomena also covers transport in electrolyte
solutions, thanks to a dedicated chapter in the widely used Deen
textbook~\cite{deen2012}.

A central component of these modern inclusions is the transport of
charged species. Electrolytes do not behave like the neutral solutes
encountered in standard convection, reaction, and diffusion problems;
their fluxes couple through the electric field and through the
electroneutrality constraint, and the equations that govern them carry
both familiar terms and new ones~\cite{bard2022,newman2004,deen2012}. In
our experience, students understand the Nernst--Planck equations, but
their coupling with the Poisson equation is less intuitive.

In conversations with colleagues teaching electrochemical engineering and
graduate transport phenomena at multiple institutions, and through
observations during research, a recurring misconception emerges: students
(and, on occasion, instructors) treat electroneutrality and a constant
electric field as equivalent. They are not. From Poisson's equation in
one dimension one can write
\[
-\varepsilon\, \frac{\rmd^{2}\varphi}{\rmd x^{2}} = \rho_F,
\]
where $\varepsilon$ is the permittivity of the solvent, $\varphi(x)$ is
the electric potential, and $\rho_F = F\sum_i z_i c_i$ is the
free-charge density. Defining the electric field
$E = -\rmd\varphi/\rmd x$, this is equivalent to
$-\varepsilon\, \rmd E/\rmd x = \rho_F$.

Reading this expression in the standard direction, if $E$ is constant in
space, then $\rho_F = 0$ and the system is electroneutral. The reverse
implication, however, is the one most often assumed: if $\rho_F = 0$,
must $E$ therefore be constant? It is tempting to read Poisson's
equation that way, but it is not what the equation says. The argument
that justifies electroneutrality in the bulk is not a statement about
the field; it is a statement about scales. Once one non-dimensionalizes
the equation, the ratio of a Debye length to a macroscopic length
appears as a small prefactor on the left-hand side. Because that
prefactor is small, the right-hand side is forced to be small, i.e.\
$\rho_F \approx 0$; nothing about that argument restricts $E$.
Electroneutral electrolytes routinely support electric fields that vary
in space~\cite{bazant2005,jarvey2022,marcicki2014}.

The result is not new. It is implicitly known when the concept of
ambipolar diffusivity is introduced~\cite{newman2004,deen2012}, where
the electric field is tied to the concentration gradients of all the
ions. To our knowledge, however, it has not been explicitly shown and
packaged into a self-contained problem aimed at students. We propose
one here, built around the limiting-current cell of Deen,
Example~15.3-3~\cite{deen2012}. Deen's example contains the essential
physics, but in our experience it is too short and not pedagogical
enough for a student to immediately grasp. The example also stops short
of explicitly enforcing the conceptual point above, which is precisely
what makes the Nernst--Planck and Poisson equations click for students.
In the proposed problems, students set up a steady one-dimensional
Poisson--Nernst--Planck system, non-dimensionalize it, identify the
small parameter $\delta = \lambda_D/\ell$ (the ratio of the Debye length
to the half-cell length), take the outer (bulk) limit, and solve. The
resulting bulk profiles satisfy electroneutrality at every interior
point, yet the dimensionless electric field varies spatially. The
impact of the electrochemical flux on the dimensionless electric field
can also be easily obtained.

A companion home problem revisits the cell after adding background
electrolyte. Students introduce a new dimensionless ratio, the ratio of
active to background concentrations, and solve the three-species
Poisson--Nernst--Planck system in closed form. Sweeping this ratio from
background-dominated to comparable concentrations, they observe how the
electric field and the limiting current change with the ratio: a
background-dominated cell flattens the electric field and halves the
limiting current, recovering the supporting-electrolyte limit familiar
from textbooks~\cite{deen2012}. The contrast between the two cases
underscores the concept: identical governing equations, identical
outer constraint of electroneutrality, and yet very different field
structures.

Beyond the central conceptual point, the problem reinforces tools that
recur throughout the chemical engineering curriculum:
non-dimensionalization and conservation equations, among others. The
level of mathematical detail is appropriate for electrochemical
engineering and graduate transport phenomena. For graduate students,
numerical resolution of the electrical double layer near the electrodes
or inclusion of Butler--Volmer kinetics could also be
added~\cite{bazant2005,jarvey2022}.

\section*{Learning objectives}
\addcontentsline{toc}{section}{Learning objectives}

By the end of this problem, students should be able to:
\begin{enumerate}[label=(\roman*)]
  \item distinguish electroneutrality from a constant electric field, and
        recognize that electroneutrality alone does not determine the
        field;
  \item non-dimensionalize the one-dimensional Poisson--Nernst--Planck
        system, identify the Debye-length-to-cell-length ratio as the
        small parameter, and explain why the bulk (electroneutral) limit
        is appropriate for macroscopic cells;
  \item compare a binary electrolyte with one containing varying levels of
        supporting electrolyte, and interpret how the resulting profiles
        and limiting current depend on the active-to-background ratio.
\end{enumerate}

\section{Problem setup}\label{sec:setup}

\begin{figure}[t]
  \centering
  \includegraphics[width=0.55\linewidth]{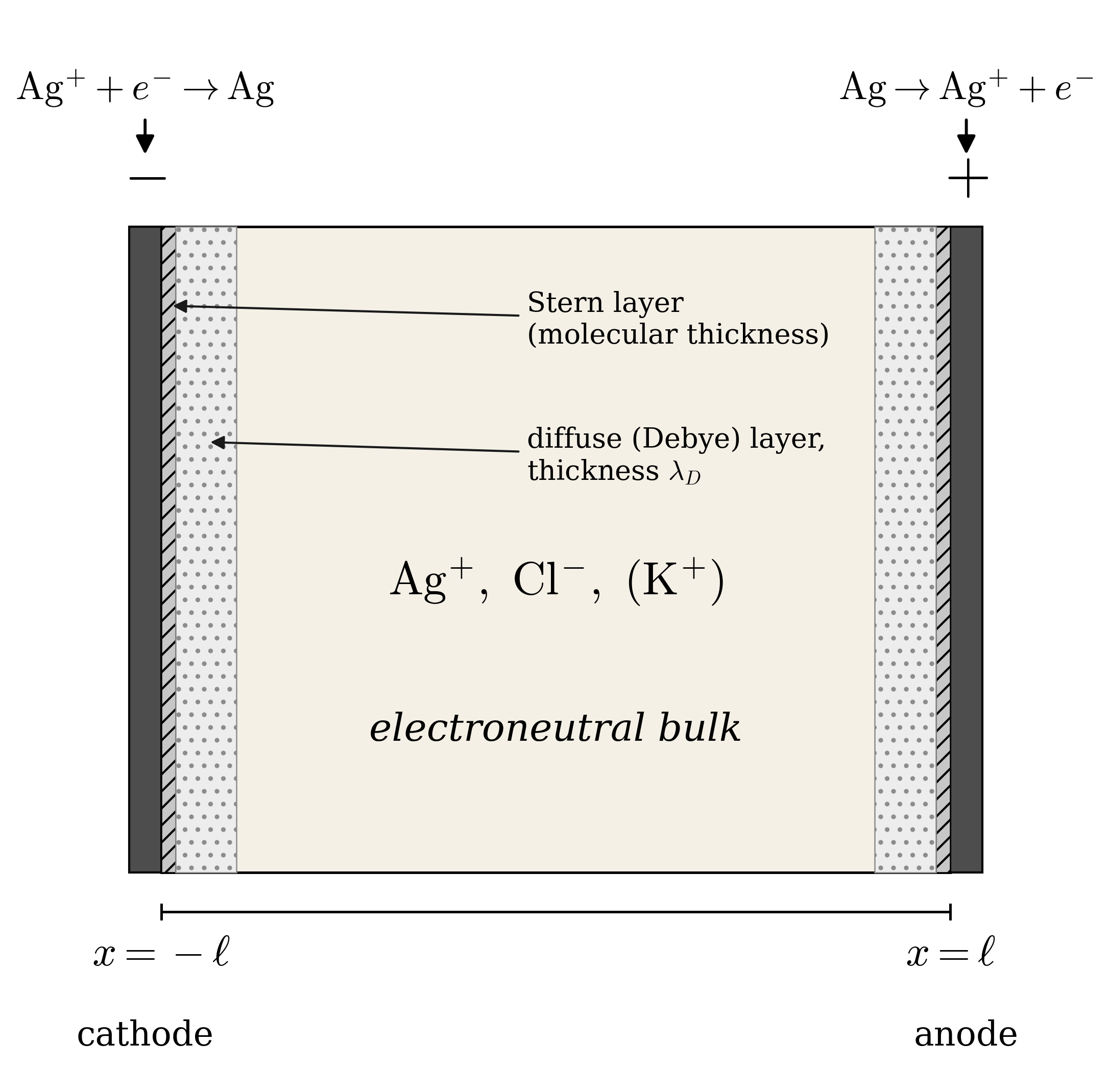}
  \caption{Idealized one-dimensional electroplating cell. Two silver
  electrodes, separated by a distance $2\ell$, bound an aqueous AgCl
  electrolyte (with optional KCl background). The cathodic deposition
  $\Ag^{+}+e^{-}\to\Ag$ occurs at $x=-\ell$; the anodic dissolution
  $\Ag\to\Ag^{+}+e^{-}$ occurs at $x=\ell$. A Stern layer of molecular
  thickness lies adjacent to each electrode, followed by a diffuse
  (Debye) layer of thickness $\lambda_D$. The effects of the Stern and
  diffuse layers are ignored in this problem.}
  \label{fig:schematic}
\end{figure}

We consider the one-dimensional electroplating cell sketched in
Figure~\ref{fig:schematic}, modeled after Deen
Example~15.3-3~\cite{deen2012}. Two silver electrodes, separated by a
distance $2\ell$, bound a quiescent electrolyte. The cell contains
either an aqueous AgCl solution alone (the class problem) or AgCl with
an added background of KCl (the home problem). An applied voltage drives
the cathodic deposition $\Ag^{+} + e^{-} \to \Ag$ at $x = -\ell$ and the
anodic dissolution $\Ag \to \Ag^{+} + e^{-}$ at $x = \ell$. The rates of
oxidation and reduction are taken equal.

Two regions, sketched in Figure~\ref{fig:schematic}, deserve note.
Adjacent to each electrode lies a Stern layer of molecular thickness,
where ions either pack densely against the surface or occupy specific
adsorption sites. Beyond the Stern layer, a diffuse (Debye) layer of
thickness $\lambda_D$ of order a few nanometers screens the electrode
charge and bridges the surface to the neutral bulk. We do not resolve
either layer in this problem, and as such $x = -\ell$ and $x = \ell$
are effectively the electrode/bulk interfaces. We treat the bulk region
as the only region of interest.

We note that this idealization conceals real physics. In a complete
treatment, the electrode kinetics (Butler--Volmer or Marcus, for
instance) couple to the diffuse-layer structure (the Frumkin
correction), and the applied voltage partitions among the Stern layer,
the diffuse layer, and the ohmic bulk~\cite{bazant2005,jarvey2022}. We
further assume that convection is absent (no stirring; the profile is
gravity-stable), the system is at steady state, and Cl$^{-}$ (and, in
the home problem, K$^{+}$) is unreactive at both electrodes, so its
average concentration is preserved as well.

We use a lowercase symbol for dimensional fluxes ($n_i$) and reserve
the capital ($N_i$) for their non-dimensional counterparts.

\section{Class problem: binary electrolyte}\label{sec:class}

We begin with a single salt in solution. Let $c_\Ag(x)$ and $c_\Cl(x)$
denote the molar concentrations of Ag$^{+}$ and Cl$^{-}$, with valences
$z_\Ag = +1$ and $z_\Cl = -1$.

\subsection*{Problem statement}

Before the cell is energized, the electrolyte has uniform concentration
$c_0$, so $c_\Ag(x) = c_\Cl(x) = c_0$. When the cell is energized, the
applied voltage drives a constant Ag$^{+}$ flux $n_\Ag = n_0$ between
the electrodes; we will treat $n_0$ as a parameter and study how the
cell responds. Cl$^{-}$ is unreactive at both electrodes.

Complete the following parts:
\begin{parts}
  \item Write the Nernst--Planck flux for each ion.
  \item Write Poisson's equation that closes the system.
  \item State the boundary conditions ($n_\Ag = n_0$ at both electrodes,
        $n_\Cl = 0$ since Cl$^{-}$ is unreactive) and the conservation
        conditions: at steady state with no leak through the electrodes,
        the mean of each species over the cell equals its initial value.
        Combine these conditions with part~(a) to write explicit balance
        equations for $c_\Ag$ and $c_\Cl$.
  \item Non-dimensionalize the system using $X = x/\ell$,
        $C_i = c_i/c_0$, $\Phi = \varphi/V_T$ with thermal voltage
        $V_T = RT/F$, and the dimensionless flux
        $N_i = n_i\ell/(D_\Ag c_0)$. Translate the boundary and
        conservation conditions.
  \item Identify $\delta = \lambda_D/\ell$ as a small parameter for
        macroscopic cells. Use bulk electroneutrality at leading order
        to set $C_\Ag = C_\Cl \equiv C$, and use the Cl$^{-}$ equation
        to eliminate $\rmd \Phi/\rmd X$ from the Ag$^{+}$ flux equation.
  \item Solve for $C(X)$ and $\rmd \Phi/\rmd X$. Plot both for several
        values of $N_0$.
  \item Decompose the Ag$^{+}$ flux into its diffusive and migrative
        parts. Repeat for Cl$^{-}$. How do the relative diffusivities of
        Ag$^{+}$ and Cl$^{-}$ enter the final result, if at all?
  \item Find the limiting current and explain what it means physically.
\end{parts}

\subsection*{Solution}

\paragraph{(a)} Each ionic flux obeys the Nernst--Planck relation,
\begin{equation}
n_i \;=\; -D_i \frac{\rmd c_i}{\rmd x}
        \;-\; \frac{D_i z_i F}{RT}\, c_i \frac{\rmd \varphi}{\rmd x},
\label{eq:np}
\end{equation}
where $D_i$ is the diffusivity of species $i$, $F$ is Faraday's
constant, $R$ is the gas constant, and $T$ is the absolute temperature.
At steady state with no homogeneous reaction,
$\rmd n_i/\rmd x = 0$, so each flux $n_i$ is constant in $x$.

\paragraph{(b)} Poisson's equation closes the system,
\begin{equation}
\frac{\rmd^2 \varphi}{\rmd x^2}
\;=\; -\frac{F}{\varepsilon}\sum_i z_i c_i,
\label{eq:poisson}
\end{equation}
where $\varepsilon$ is the permittivity of the solvent (taken constant)
and the sum runs over Ag and Cl.

\paragraph{(c)} The boundary conditions are $n_\Ag = n_0$ at both
electrodes and $n_\Cl = 0$. At steady state, the mean of each species
over the cell equals its initial value:
\begin{equation}
\frac{1}{2\ell}\int_{-\ell}^{\ell} c_i\, \rmd x
\;=\; \avg{c_i} \;=\; c_0.
\label{eq:cons}
\end{equation}
The conservation condition matters because the boundary conditions
only set the derivatives of $c_i$ (through the fluxes), not their
absolute values; Eq.~(\ref{eq:cons}) is what fixes the integration
constants. Substituting $n_\Ag = n_0$ and $n_\Cl = 0$ into
Eq.~(\ref{eq:np}) gives explicit balance equations,
\begin{align}
-D_\Ag \frac{\rmd c_\Ag}{\rmd x}
  - \frac{D_\Ag F}{RT}\, c_\Ag \frac{\rmd \varphi}{\rmd x} &\;=\; n_0,
  \label{eq:bin-Ag}\\
-D_\Cl \frac{\rmd c_\Cl}{\rmd x}
  + \frac{D_\Cl F}{RT}\, c_\Cl \frac{\rmd \varphi}{\rmd x} &\;=\; 0.
  \label{eq:bin-Cl}
\end{align}

\paragraph{(d)} Scale lengths by $\ell$, concentrations by $c_0$, the
potential by the thermal voltage $V_T = RT/F$ ($\approx 25.7\,$mV at
$25\,^\circ$C), and the flux of species $i$ by $D_\Ag c_0/\ell$:
\begin{equation}
X = \frac{x}{\ell},\quad
C_i = \frac{c_i}{c_0},\quad
\Phi = \frac{\varphi}{V_T},\quad
N_i = \frac{n_i \ell}{D_\Ag c_0}.
\label{eq:nondim}
\end{equation}
The boundary conditions become $N_\Ag = N_0$ and $N_\Cl = 0$, where
$N_0 = n_0\ell/(D_\Ag c_0)$ is the dimensionless imposed flux.
Equations~(\ref{eq:bin-Ag})--(\ref{eq:bin-Cl}) become
\begin{align}
-\frac{\rmd C_\Ag}{\rmd X} - C_\Ag \frac{\rmd \Phi}{\rmd X} &\;=\; N_0,
  \label{eq:bin-Ag-nd}\\
\hat{D}_\Cl\!\left(-\frac{\rmd C_\Cl}{\rmd X}
  + C_\Cl \frac{\rmd \Phi}{\rmd X}\right) &\;=\; 0,\qquad
  \hat{D}_i = \frac{D_i}{D_\Ag},
  \label{eq:bin-Cl-nd}
\end{align}
with $\hat{D}_\Ag = 1$. Equation~(\ref{eq:poisson}) becomes
\begin{equation}
2\delta^2 \frac{\rmd^2 \Phi}{\rmd X^2}
\;=\; -\sum_i z_i C_i,\qquad
\delta = \frac{\lambda_D}{\ell},\qquad
\lambda_D = \sqrt{\frac{\varepsilon V_T}{2 F c_0}},
\label{eq:poisson-nd}
\end{equation}
and the conservation constraint Eq.~(\ref{eq:cons}) reads
$\tfrac{1}{2}\!\int_{-1}^{1} C_i\, \rmd X = 1$ for $i\in\{\Ag,\Cl\}$.
The Debye-to-half-cell ratio $\delta = \lambda_D/\ell$ measures how far
the electric field penetrates from a charged surface relative to the
cell size; for $c_0 \sim 1\,$mM in water at $25\,^\circ$C and
$\ell \sim 1\,$mm, $\lambda_D \sim 10\,$nm and $\delta \sim 10^{-5}$.
We treat $\delta$ as the small parameter.

\paragraph{(e)} Because $\delta \ll 1$, expand the dependent variables
in regular perturbation series in $\delta^2$:
\[
C_i(X) = C_i^{(0)}(X) + \delta^2 C_i^{(1)}(X) + \cdots,\qquad
\Phi(X) = \Phi^{(0)}(X) + \delta^2 \Phi^{(1)}(X) + \cdots.
\]
Substituting into Eq.~(\ref{eq:poisson-nd}), the left side is order
$\delta^2$ while the right side is order one. Matching at leading order
forces bulk electroneutrality,
\begin{equation}
\sum_i z_i C_i^{(0)} \;=\; 0.
\label{eq:neut}
\end{equation}
This is the central conceptual moment: when $\delta = \lambda_D/\ell$
is small, the prefactor $2\delta^2$ on the left-hand side of
Eq.~(\ref{eq:poisson-nd}) is essentially zero, so the only way for the
equation to balance is for the right-hand side to vanish. Importantly,
the argument never invoked a constant field: nothing in
Eq.~(\ref{eq:poisson-nd}) or Eq.~(\ref{eq:neut}) restricts
$\rmd \Phi/\rmd X$. Electroneutrality holds because $\lambda_D/\ell$ is
a small number, \emph{not} because the electric field is constant.

We work at leading order from here on and drop the superscript~(0).
For the binary system, Eq.~(\ref{eq:neut}) gives $C_\Ag = C_\Cl \equiv C$.
The Cl$^{-}$ balance Eq.~(\ref{eq:bin-Cl-nd}) with the common $C$ then
yields
\begin{equation}
-\frac{\rmd C}{\rmd X} + C \frac{\rmd \Phi}{\rmd X} = 0,
\qquad
\frac{\rmd \Phi}{\rmd X} = \frac{1}{C}\frac{\rmd C}{\rmd X}.
\label{eq:bin-dphi}
\end{equation}
Substituting Eq.~(\ref{eq:bin-dphi}) into Eq.~(\ref{eq:bin-Ag-nd}) with
$C_\Ag = C$ gives $-2\,\rmd C/\rmd X = N_0$.

\begin{figure}[t]
  \centering
  \includegraphics[width=0.95\linewidth]{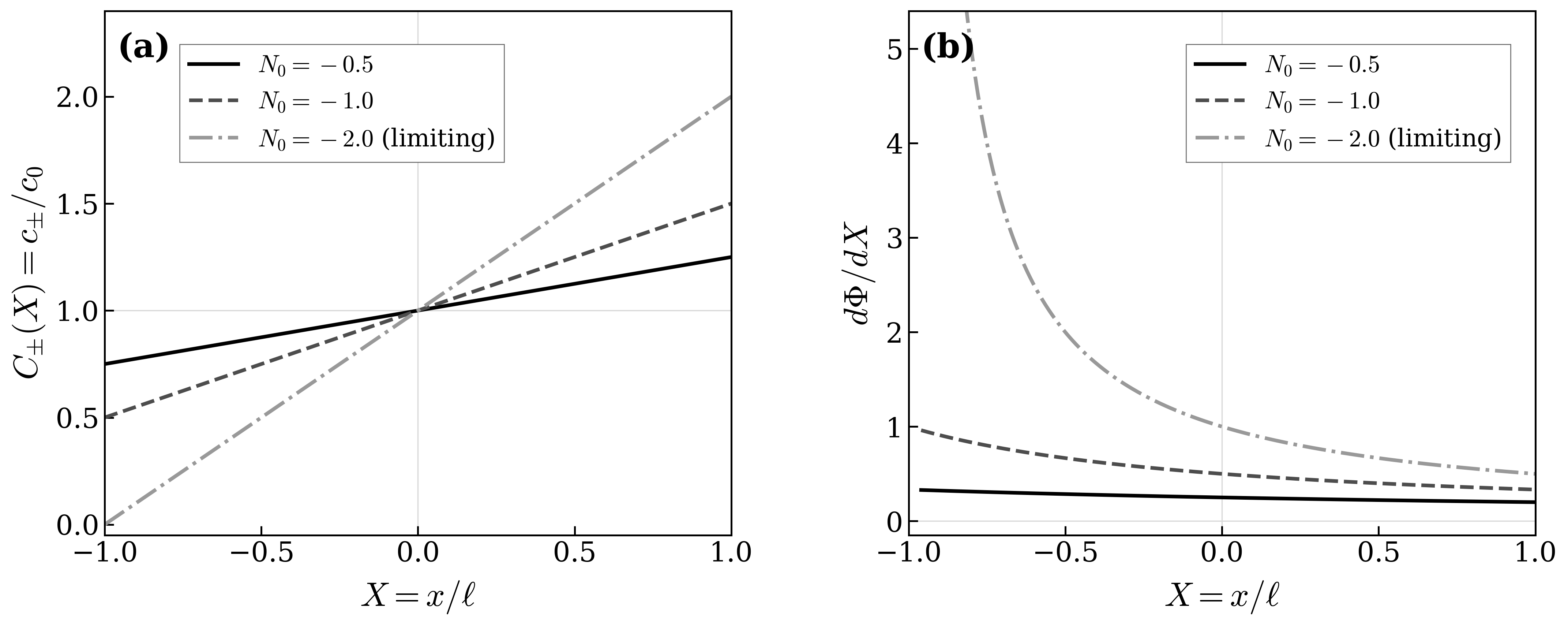}
  \caption{Bulk solution of the binary class problem at three values of
  the dimensionless cation flux. (a) Cation concentration
  $C_{\pm}(X) = c_{\pm}/c_0$, linear in $X$, reaches zero at the cathode
  for the limiting flux $N_0 = -2$. (b) Dimensionless field gradient
  $\rmd \Phi/\rmd X$ is positive and varies smoothly across the cell
  for sub-limiting fluxes; at the limiting condition it diverges near
  the cathode, even though the bulk is electroneutral at every interior
  point.}
  \label{fig:bin}
\end{figure}

\paragraph{(f)} Integrating with the conservation constraint
$\avg{C}=1$ on the symmetric domain (so $C(0)=1$),
\begin{equation}
C(X) \;=\; 1 - \frac{N_0\, X}{2},
\label{eq:bin-C}
\end{equation}
and Eq.~(\ref{eq:bin-dphi}) gives the dimensionless field
\begin{equation}
\frac{\rmd \Phi}{\rmd X} \;=\; -\,\frac{N_0}{2 - N_0\, X}.
\label{eq:bin-field}
\end{equation}
Figure~\ref{fig:bin} plots $C(X)$ and $\rmd\Phi/\rmd X$ for
$N_0\in\{-0.5,-1,-2\}$. Cation and anion concentrations are equal at
every interior point (electroneutrality), yet the field gradient is
monotonically varying. The cathode-side concentration
$C(-1) = 1 + N_0/2$ reaches zero at the limiting flux $N_0 = -2$, and
the field diverges as $1/(1+X)$ near the cathode.

\paragraph{(g)} Splitting the Ag$^{+}$ flux into its diffusive and
migrative parts (using $\hat{D}_\Ag = 1$ and
$\rmd C/\rmd X = -N_0/2$ from Eq.~(\ref{eq:bin-C})),
\begin{equation}
N_\Ag^{\mathrm{diff}}
  = -\frac{\rmd C_\Ag}{\rmd X}
  = \frac{N_0}{2},\qquad
N_\Ag^{\mathrm{mig}}
  = -C_\Ag \frac{\rmd \Phi}{\rmd X}
  = \frac{N_0}{2}.
\label{eq:bin-Ag-split}
\end{equation}
The two halves contribute equally and add to $N_0$. For Cl$^{-}$,
\begin{equation}
N_\Cl^{\mathrm{diff}}
  = -\hat{D}_\Cl\frac{\rmd C_\Cl}{\rmd X}
  = \frac{\hat{D}_\Cl N_0}{2},\qquad
N_\Cl^{\mathrm{mig}}
  = +\hat{D}_\Cl C_\Cl \frac{\rmd \Phi}{\rmd X}
  = -\frac{\hat{D}_\Cl N_0}{2},
\label{eq:bin-Cl-split}
\end{equation}
which cancel: $N_\Cl = 0$ throughout the bulk, recovering the boundary
condition. The diffusivity ratio $\hat{D}_\Cl$ enters as a common
factor and drops out of the result; only $D_\Ag$ sets the scale of
the Ag$^{+}$ flux.

\paragraph{(h)} From Eq.~(\ref{eq:bin-C}), the cathode-side
concentration is $C(-1) = 1 + N_0/2$. Reducing the Ag$^{+}$ flux means
making $N_0$ more negative, which lowers $C(-1)$ until it hits zero at
$N_0 = -2$. At that point the cathode has run out of cations: no
further increase in flux is possible because Ag$^{+}$ would have to be
supplied at a concentration that does not exist. This is the limiting
current. The dimensional limiting current density is
\begin{equation}
i_L \;=\; F n_\Ag \;=\; -\,\frac{2 F D_\Ag c_0}{\ell},
\label{eq:bin-iL}
\end{equation}
in agreement with Deen Eq.~(15.3-14)~\cite{deen2012} (note that the
cell length there is $2\ell$). Only $D_\Ag$ enters the limiting
current, a feature of binary electrolytes with monovalent cation and
anion~\cite{newman2004}.

\section{Home problem: added background electrolyte}\label{sec:home}

We now repeat the analysis for the same cell with KCl added as a
background electrolyte. Three species (Ag$^{+}$, Cl$^{-}$, K$^{+}$)
move in the cell with valences $z_\Ag = z_\Kp = +1$ and
$z_\Cl = -1$.

\subsection*{Problem statement}

Before the cell is energized, the AgCl average concentration is $c_0$
and the KCl average concentration is $c_b$ (a positive constant whose
value we will treat as a parameter). When the cell is energized,
$n_\Ag = n_0$ again at both electrodes; both Cl$^{-}$ and K$^{+}$
remain unreactive ($n_\Cl = n_\Kp = 0$). At steady state with no leak
through the electrodes, the cell-mean concentrations are preserved:
$\avg{c_\Ag} = c_0$, $\avg{c_\Kp} = c_b$, and
$\avg{c_\Cl} = c_0 + c_b$.

Complete the following parts:
\begin{parts}
  \item Write the Nernst--Planck flux for each of the three species
        (Ag$^{+}$, Cl$^{-}$, K$^{+}$).
  \item State the boundary conditions: $n_\Ag = n_0$ and
        $n_\Cl = n_\Kp = 0$.
  \item Combine (a) and (b) to write three explicit balance equations
        for $c_\Ag$, $c_\Kp$, and $c_\Cl$.
  \item Working in the non-dimensional variables $X = x/\ell$,
        $C_i = c_i/c_0$, $\Phi = \varphi/V_T$,
        $N_i = n_i\ell/(D_\Ag c_0)$, and the active-to-background
        ratio $\Gamma = c_0/c_b$, add the three balance equations and
        use bulk electroneutrality to eliminate the migration term.
        Integrate, using the conservation conditions, and obtain
        closed-form expressions for $C_\Cl(X)$, the dimensionless
        field $\rmd\Phi/\rmd X$, and the cation profile $C_\Ag(X)$ as
        functions of $X$, $N_0$, and $\Gamma$.
  \item Evaluate $C_\Ag(X)$ and $\rmd\Phi/\rmd X$ numerically at
        $N_0 = -0.5$ for $\Gamma\in\{0.01, 0.1, 0.5, 1\}$ and plot
        them.
  \item Solve numerically for the limiting flux at each of
        $\Gamma\in\{0.01, 0.1, 0.5, 1\}$, and discuss how the limiting
        current interpolates between the supporting-electrolyte limit
        and the binary class-problem result.
  \item \textbf{Bonus.} Solve the limiting condition for
        $N_0^{\mathrm{lim}}$ numerically over a wider range of $\Gamma$
        (e.g., $10^{-3} \leq \Gamma \leq 10^{3}$) and plot
        $|N_0^{\mathrm{lim}}|$ versus $\Gamma$ on a logarithmic
        $\Gamma$ axis. Verify the asymptotes
        $|N_0^{\mathrm{lim}}| \to 1$ as $\Gamma \to 0$
        (supporting-electrolyte limit) and
        $|N_0^{\mathrm{lim}}| \to 2$ as $\Gamma \to \infty$ (binary
        limit), and read off the four discrete values from part~(f).
\end{parts}

\subsection*{Solution}

\paragraph{(a)} Each of the three species obeys Eq.~(\ref{eq:np}) with
its own valence: Ag$^{+}$ ($z_\Ag = +1$), Cl$^{-}$ ($z_\Cl = -1$), and
K$^{+}$ ($z_\Kp = +1$). At steady state with no homogeneous reaction,
every flux is independent of $x$.

\paragraph{(b)} The boundary conditions at both electrodes are
$n_\Ag = n_0$ (specified Ag$^{+}$ flux) and $n_\Cl = n_\Kp = 0$
(Cl$^{-}$ and K$^{+}$ are unreactive).

\paragraph{(c)} Substituting the boundary conditions into
Eq.~(\ref{eq:np}) and dividing each balance by the diffusivity of that
species gives three first-order ODEs for the dimensional
concentrations:
\begin{align}
\frac{\rmd c_\Ag}{\rmd x}
  + \frac{F}{RT}\, c_\Ag \frac{\rmd \varphi}{\rmd x}
  &\;=\; -\frac{n_0}{D_\Ag},
  \label{eq:hp-Ag}\\
\frac{\rmd c_\Kp}{\rmd x}
  + \frac{F}{RT}\, c_\Kp \frac{\rmd \varphi}{\rmd x}
  &\;=\; 0,
  \label{eq:hp-K}\\
\frac{\rmd c_\Cl}{\rmd x}
  - \frac{F}{RT}\, c_\Cl \frac{\rmd \varphi}{\rmd x}
  &\;=\; 0.
  \label{eq:hp-Cl}
\end{align}
Equations~(\ref{eq:hp-Ag})--(\ref{eq:hp-Cl}) are the analog of
Eqs.~(\ref{eq:bin-Ag})--(\ref{eq:bin-Cl}) of the class problem,
augmented with a third species.

\paragraph{(d)} Non-dimensionalizing with the same scales as in the
class problem (Eq.~(\ref{eq:nondim})) and the active-to-background
ratio
\begin{equation}
\Gamma \;=\; \frac{c_0}{c_b},
\label{eq:gamma}
\end{equation}
Eqs.~(\ref{eq:hp-Ag})--(\ref{eq:hp-Cl}) become
\begin{align}
\frac{\rmd C_\Ag}{\rmd X}
  + C_\Ag \frac{\rmd \Phi}{\rmd X}
  &\;=\; -N_0,
  \label{eq:hp-Ag-nd}\\
\frac{\rmd C_\Kp}{\rmd X}
  + C_\Kp \frac{\rmd \Phi}{\rmd X}
  &\;=\; 0,
  \label{eq:hp-K-nd}\\
\frac{\rmd C_\Cl}{\rmd X}
  - C_\Cl \frac{\rmd \Phi}{\rmd X}
  &\;=\; 0.
  \label{eq:hp-Cl-nd}
\end{align}
Adding Eqs.~(\ref{eq:hp-Ag-nd})--(\ref{eq:hp-Cl-nd}),
\[
\frac{\rmd}{\rmd X}\!\bigl(C_\Ag + C_\Kp + C_\Cl\bigr)
+ (C_\Ag + C_\Kp - C_\Cl)\frac{\rmd \Phi}{\rmd X}
= -N_0.
\]
Bulk electroneutrality (Eq.~(\ref{eq:neut}), restated for the
three-species system) gives $C_\Ag + C_\Kp - C_\Cl = 0$, so the
migration term vanishes and the sum satisfies
$\rmd(C_\Ag + C_\Kp + C_\Cl)/\rmd X = -N_0$. Integrating and using
$\avg{C_\Ag}=1$, $\avg{C_\Kp}=1/\Gamma$,
$\avg{C_\Cl}=1+1/\Gamma$ to fix the integration constant,
\begin{equation}
C_\Ag + C_\Kp + C_\Cl \;=\; 2\!\left(1+\tfrac{1}{\Gamma}\right) - N_0\, X.
\label{eq:hp-sum}
\end{equation}
Combining Eq.~(\ref{eq:hp-sum}) with electroneutrality
($C_\Ag + C_\Kp = C_\Cl$) yields the chloride profile
\begin{equation}
C_\Cl(X) \;=\; \left(1 + \tfrac{1}{\Gamma}\right) - \frac{N_0}{2} X.
\label{eq:hp-CCl}
\end{equation}
The dimensionless field follows from the Cl$^{-}$ balance
Eq.~(\ref{eq:hp-Cl-nd}) as
$\rmd \Phi/\rmd X = (1/C_\Cl)\,\rmd C_\Cl/\rmd X$, which evaluates
with Eq.~(\ref{eq:hp-CCl}) to
\begin{equation}
\frac{\rmd \Phi}{\rmd X}
\;=\; -\,\frac{N_0}{2\!\left(1+\tfrac{1}{\Gamma}\right) - N_0\, X},
\label{eq:hp-field}
\end{equation}
exact for all $\Gamma$. For $\Gamma\ll 1$, the denominator is
dominated by $2/\Gamma$ and $\rmd\Phi/\rmd X \approx -N_0\Gamma/2$ is
small (the supporting-electrolyte limit). For $\Gamma\to\infty$, the
$1/\Gamma$ term drops out and Eq.~(\ref{eq:hp-field}) recovers
Eq.~(\ref{eq:bin-field}) of the class problem.

With $\rmd \Phi/\rmd X$ now known, the K$^{+}$ balance
Eq.~(\ref{eq:hp-K-nd}) becomes a first-order ODE for $C_\Kp$ alone.
Substituting $\rmd \Phi/\rmd X = (1/C_\Cl)\,\rmd C_\Cl/\rmd X$ recasts
Eq.~(\ref{eq:hp-K-nd}) as $\rmd(C_\Kp C_\Cl)/\rmd X = 0$, so
$C_\Kp C_\Cl$ is constant in $X$. Equivalently, $C_\Kp(X)$ is
inversely proportional to $C_\Cl(X)$. The proportionality constant is
fixed by $\avg{C_\Kp}=1/\Gamma$; the average of $1/C_\Cl$ over the
linear Cl$^{-}$ profile (Eq.~(\ref{eq:hp-CCl})) evaluates to a
logarithm, giving
\begin{equation}
C_\Kp(X)
\;=\; \frac{-N_0}{\Gamma\, C_\Cl(X)\,\ln\!\left(\dfrac{C_\Cl(1)}{C_\Cl(-1)}\right)},
\label{eq:hp-CK}
\end{equation}
with $C_\Cl(\pm 1)$ read from Eq.~(\ref{eq:hp-CCl}). Finally, the
cation profile follows from electroneutrality:
\begin{equation}
C_\Ag(X) \;=\; C_\Cl(X) - C_\Kp(X).
\label{eq:hp-CAg}
\end{equation}

\begin{figure}[t]
  \centering
  \includegraphics[width=0.98\linewidth]{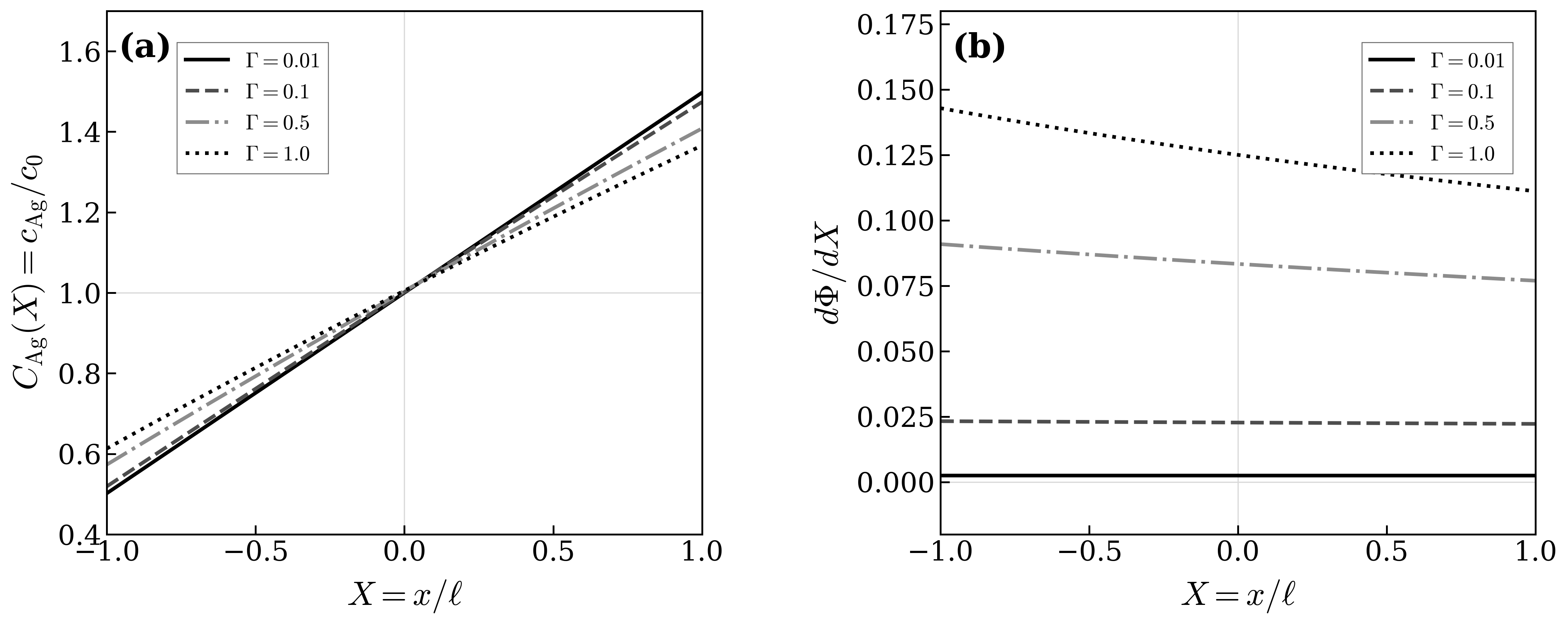}
  \caption{Home-problem profiles at fixed $N_0 = -0.5$ for four values
  of the active-to-background ratio
  $\Gamma\in\{0.01, 0.1, 0.5, 1\}$. (a) Cation concentration
  $C_\Ag(X)$. At small $\Gamma$ (background dominant) the profile is
  nearly linear with slope $-N_0$; at $\Gamma$ of order one, the
  profile picks up the curvature of the binary class problem.
  (b) Field gradient $\rmd \Phi/\rmd X$ ranges from approximately
  $-N_0\Gamma/2$ at small $\Gamma$ (supporting limit, nearly zero) up
  to a substantial value at $\Gamma = 1$.}
  \label{fig:home}
\end{figure}

\paragraph{(e)} We evaluate the cation profile (Eq.~(\ref{eq:hp-CAg}))
and the dimensionless field (Eq.~(\ref{eq:hp-field})) at $N_0 = -0.5$
for $\Gamma\in\{0.01, 0.1, 0.5, 1\}$. Figure~\ref{fig:home} displays
the results.

\paragraph{(f)} The limiting current is reached when the cathode-side
cation concentration vanishes, $C_\Ag(-1) = 0$.
Equation~(\ref{eq:hp-CAg}) then requires $C_\Cl(-1) = C_\Kp(-1)$.
Substituting Eq.~(\ref{eq:hp-CK}) at $X = -1$ with $N_0$ set to its
limiting value $N_0^{\mathrm{lim}}$ (negative for cathodic flux, so
$-N_0^{\mathrm{lim}} = |N_0^{\mathrm{lim}}|$) yields
\begin{equation}
\bigl[C_\Cl(-1)\bigr]^{2}
\;=\;
\frac{|N_0^{\mathrm{lim}}|}{\Gamma\,\ln\!\left(\dfrac{C_\Cl(1)}{C_\Cl(-1)}\right)},
\label{eq:hp-limit}
\end{equation}
which determines the limiting non-dimensional flux
$N_0^{\mathrm{lim}}$ at any given $\Gamma$. Solving
Eq.~(\ref{eq:hp-limit}) numerically (e.g., bisection on
$-2 < N_0^{\mathrm{lim}} < 0$, since $N_0^{\mathrm{lim}} \to -1$ in
the supported limit and $\to -2$ in the binary limit) gives
$N_0^{\mathrm{lim}} \approx -1.003,\, -1.032,\, -1.130,\, -1.217$ at
$\Gamma = 0.01,\, 0.1,\, 0.5,\, 1$, respectively. The corresponding
dimensional limiting current density is
\begin{equation}
i_L \;=\; \frac{F\, N_0^{\mathrm{lim}}\, D_\Ag\, c_0}{\ell}.
\label{eq:hp-iL}
\end{equation}
As $\Gamma \to 0$ (large background), $N_0^{\mathrm{lim}} \to -1$ and
the limiting current approaches half of the binary class-problem value
(Eq.~(\ref{eq:bin-iL})); this is the supporting-electrolyte limit of
Deen~\cite{deen2012}. As $\Gamma \to \infty$ (no background),
$N_0^{\mathrm{lim}} \to -2$, recovering the binary cell. At
intermediate $\Gamma$, the limiting current interpolates smoothly
between these two extremes; even at $\Gamma = 1$, where Cl$^{-}$ from
KCl is comparable to that from AgCl,
$N_0^{\mathrm{lim}} \approx -1.22$ sits closer to the
supporting-electrolyte value ($-1$) than to the binary one ($-2$),
reflecting how strongly K$^{+}$ migration alone shifts the cell off
its binary behavior.

\begin{figure}[t]
  \centering
  \includegraphics[width=0.7\linewidth]{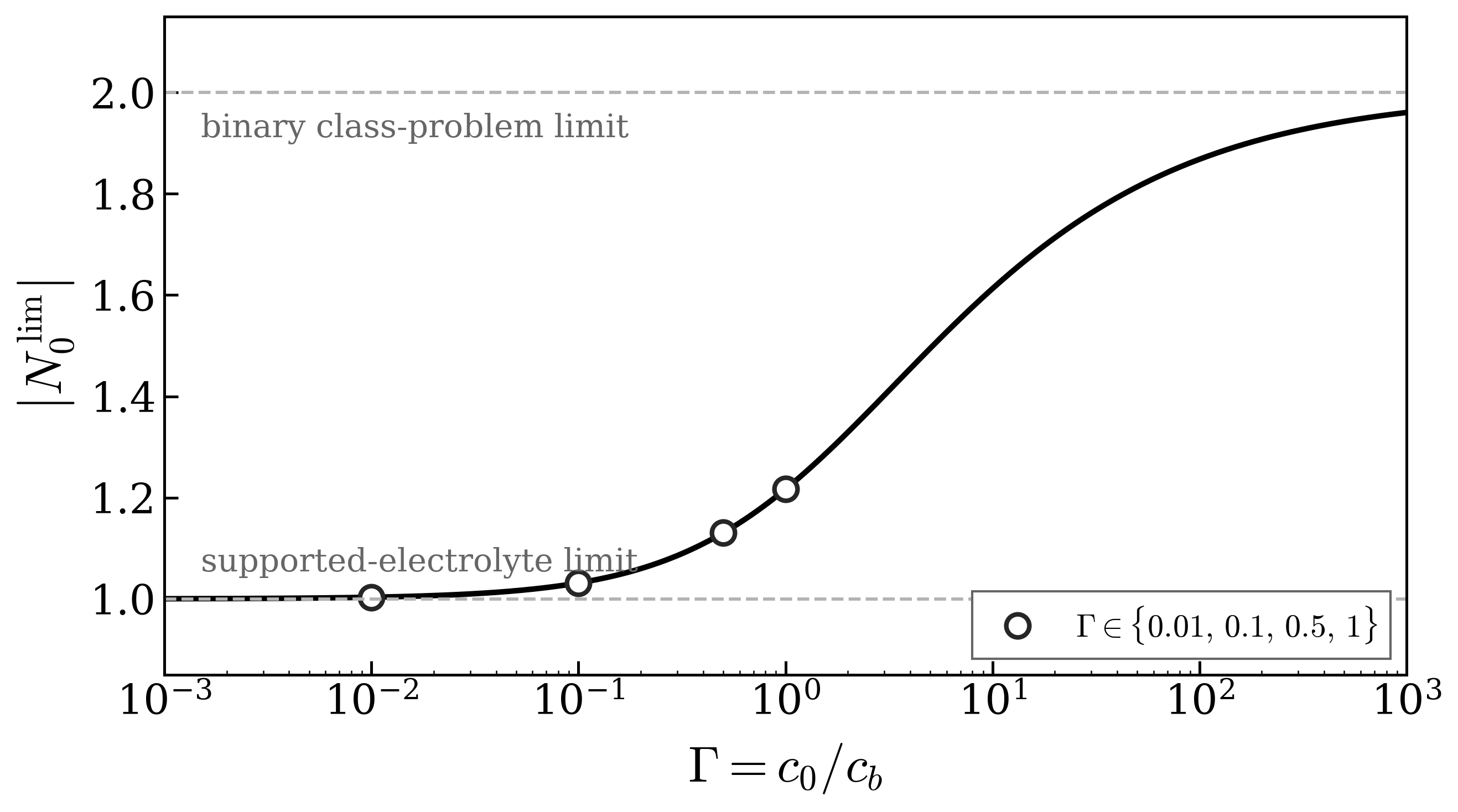}
  \caption{Magnitude of the dimensionless limiting flux
  $|N_0^{\mathrm{lim}}|$ versus the active-to-background ratio
  $\Gamma$ on a logarithmic $\Gamma$ axis, obtained by bisecting
  Eq.~(\ref{eq:hp-limit}). The curve interpolates monotonically from
  the supporting-electrolyte limit ($|N_0^{\mathrm{lim}}|=1$, dashed
  line near the bottom) to the binary class-problem limit
  ($|N_0^{\mathrm{lim}}|=2$, dashed line near the top). Open circles
  mark the four $\Gamma$ values used in part~(f).}
  \label{fig:limit}
\end{figure}

\paragraph{(g)} Solving Eq.~(\ref{eq:hp-limit}) by bisection over a
wider range of $\Gamma$ produces Figure~\ref{fig:limit}: a smooth,
monotonic curve that saturates at $|N_0^{\mathrm{lim}}|=1$ as
$\Gamma\to 0$ and approaches $|N_0^{\mathrm{lim}}|=2$ as
$\Gamma\to\infty$. The four open circles mark the discrete values
from part~(f).

\section{Concluding remarks}

This class and home problem provides students with a concrete
demonstration that electroneutrality is necessary, but not sufficient,
for a constant electric field. From scaling and a perturbation
analysis of the one-dimensional Poisson--Nernst--Planck system,
students obtain the bulk solution in two complementary settings: a
binary AgCl cell, where the dimensionless field $\rmd\Phi/\rmd X$
varies as $-N_0/(2 - N_0 X)$ and diverges at the cathode under
limiting current; and a cell with added background electrolyte, in
which the same system is solved in closed form for any
active-to-background ratio $\Gamma$. Numerical evaluation at
$\Gamma\in\{0.01, 0.1, 0.5, 1\}$ shows that the dimensionless field
shrinks as $\Gamma$ decreases (vanishing in the supporting-electrolyte
limit), and the limiting current interpolates from half of the binary
value at small $\Gamma$ to the full binary value as
$\Gamma\to\infty$. Together, the two problems reinforce the
pedagogical point: a constant field implies electroneutrality, but
electroneutrality alone does not pin the field.

The mathematical level fits an electrochemical engineering elective
and graduate transport phenomena. Extensions for graduate students
include numerical resolution of the inner Stern and diffuse layers
via matched asymptotics~\cite{bazant2005} and incorporating
Butler--Volmer kinetics~\cite{jarvey2022}.

\section*{Statement on AI use}
\addcontentsline{toc}{section}{Statement on AI use}

A generative AI assistant (Claude, Anthropic) was used to help draft
portions of the manuscript and to produce figures. The conceptual
content, problem statements, and derivations are the author's.

\bibliographystyle{ieeetr}
\bibliography{refs}

\end{document}